\documentclass[a4paper,11pt]{article}
\pdfoutput=1 
\usepackage{jcappub}
\usepackage{float}
\usepackage{color}
\usepackage{xcolor}
\usepackage{booktabs}
\usepackage{amsmath}
\usepackage{wrapfig}

\newcommand{\fix}[1]{\textbf{#1}}
\usepackage{graphicx}
\usepackage{bm}
\usepackage{epstopdf}

\usepackage{subfigure}
\bibpunct{(}{)}{;}{a}{}{,}
\usepackage{subfigure}
\usepackage{amsmath}
\usepackage{amsfonts}
\usepackage{amssymb}
\usepackage{ulem} 
\usepackage{graphicx}
\usepackage{natbib}

\title{Nova Neutrinos in the Multi-Messenger Era}

\author[1]{Dafne Guetta,}
\author[1]{Yael Hillman,}
\author[1,2]{Massimo Della Valle}
\affiliation[1]{Ariel University, Ariel, Israel}
\affiliation[2]{Osservatorio Astrofisico di Capodimonte, INAF-Napoli, Italy}
\emailAdd{dafneg@ariel.ac.il}

\abstract
 {The recently discovered high energy emission from the recurrent nova RS Ophiuchi by Fermi-LAT ($>$ 100 MeV), H.E.S.S. and MAGIC ($>$ 100 GeV), hints towards a possible hadronic origin of this radiation component. 
 From the observed high energy photon flux we derive the expected number of neutrino events that could be detected by present and future neutrino telescopes in the different energy ranges. 
 Preliminary estimates indicate that with the ”next- generation” instrument IceCube-Gen2, the expected number of neutrino detections from Galactic novae, admittedly with large uncertainty, should not exceed 1 event per decade. }

\keywords{(stars:) binaries: symbiotic --- (stars:) novae, cataclysmic variables
 ---  neutrinos --- gamma rays: general}

\notoc

\begin{document}
\maketitle
\flushbottom
\newpage

\newcommand\T{\rule{0pt}{2.6ex}}       
\newcommand\B{\rule[-1.2ex]{0pt}{0pt}} 

\def\apj{ApJ}                 
\def\apjls{ApJLS}               
\def\apjl{ApJL}               
\def\apjs{ApJS}               
\def\cqg{CQG} 
\def\mnras{MNRAS}             
\def\aap{A\&A}                
\def\aaps{A\&AS}              
\def\aj{AJ}                   
\def\physrep{Phys.~Rep.}      
\def\nat{Nature}              
\def\araa{ARA\&A}             
\def\pasj{PASJ}               
\def\prd{Phys. Rev. D}        
\def\prl{Phys. Rev. Lett.}    
\def\jcap{Journal of Cosmology and Astroparticle Physics}
\def\aapr{A\&A review}                
\def\apss{Ap\&SS}                

\section{Introduction}\label{sec:intro}

A nova is a powerful eruption following a thermonuclear runaway (TNR) that occurs below the surface of a white dwarf (WD) \cite[]{Starrfield1972,Shara1981,Starrfield2008}. The TNR is the inevitable result of a critical amount of (mostly) hydrogen being pulled away from its companion, less evolved star, and accumulating on the degenerate surface of the WD. As this mass piles up in a degenerate environment, the pressure below the surface increases, causing the temperature to rise until becoming sufficiently high to ignite the hydrogen, entailing fusion in a runaway process and the violent ejection of the envelope \cite[e.g.,][]{Shara1981}, exhibited as an enormous optical brightening \cite[]{Pay1957} of order $\sim10^{4-5}$ times the solar luminosity ---  the nova eruption \cite[]{Hellier2001,Warner2003,Knigge2011}. Novae are usually discovered, following the eruption, in the optical band  but it is hardly the only range in which a nova may be observed.
Over the course of a nova cycle --- accretion, eruption and decline --- a nova producing system could possibly be observed in the infrared (IR), ultraviolet (UV), soft and hard X-rays and even $\gamma$-rays \cite[]{MacDonald1985,Itoh1990,Orio1994,Orio2009,Schaefer2010,Hillman2014,DellaValle2020,Chomiuk2021,Konig2022}. Each band, if observed, may provide clues as to the nature of the eruption, the system and its unique behavior that could help distinguish one system from the next.
However, the capacity and technical capabilities of capturing observations in the different bands are not the same for all bands, resulting in mostly optical records. While the optical rise indicates the expansion of the WD's envelope and ejection of the mass \cite[e.g.,][]{Prialnik1986}, $\gamma$-rays, if observed, are detected only days after the optical peak \cite[]{Sokolovsky2022}, implying that they must be originating from somewhere other than the WD's surface \cite[]{Metzger2015,Martin2018}.

In the past decade $\gamma$-rays were detected in a handful of systems emitting at energies higher than $100 $ MeV using the Fermi-Large Area Telescope (Fermi-LAT) \cite[]{Razzaque2010,Ackermann2014,Cheung2016,Martin2018}. \cite{Ackermann2014} investigated the likelihood of the $\gamma$-rays originating from both hadronic and leptonic processes, for the symbiotic nova (SymN) V407 Cyg and the three classical novae (CN) V1324 Sco, V959 Mon and V339 Del, but did not come to a firm conclusion regarding which emitting process is more likely to be the source.  \cite{Cheung2016} explored detected $\gamma$-rays for an additional two CNe --- V1369 Cen and V5667 Sgr \cite[]{Li2016,Li2017}, and interpreted that this high energy emission is due to particles accelerated up to $\sim$ 100 GeV at the reverse shock and undergoing hadronic interactions in the dense cooling layer downstream of the shock \cite[]{Martin2018}. 

Recently, the MAGIC (Major Atmospheric Gamma Imaging Cherenkov) and the H.E.S.S. (High Energy Stereoscopic System) telescopes have detected $\gamma$-rays of energies higher than $100$ GeV from the 2021 outburst of RS Oph --- a recurrent nova (RN) in a symbiotic system (SymRN) that erupts every $\sim15$ years \cite[]{Acciari2022,Hess2022}. When a nova eruption occurs in a symbiotic system, the ejected mass will inevitably collide with the dense wind of the red giant (RG) companion giving rise to a shock. The gas through which the shock propagates is shocked and a fraction of  particles are accelerated. The accelerated particles then emit high energy radiation in the $\gamma$-ray range such as seen in RS Oph \cite[]{Hess2022} and V407 Cyg \cite[]{Abdo2010,Martin2018}.
Another scenario that has been recently considered (\cite{Diesing2022}), is the possibility to have multiple shocks in order to reproduce the observed $\gamma$-ray spectrum.

Systems with a red dwarf (RD) donor (i.e., a cataclysmic variable (CV)) might also produce shocks in the event that the ejected mass is not expelled in a unified manner, but rather in stages or in clumps with different velocities, thus a fast clump of mass could collide with a previously ejected slower moving clump of mass. It is also plausible that the ejected mass shell may interact with an expanding mass shell that was ejected in a previous nova eruption, provided the recurrence period is short enough and enough mass was ejected. However, it is not established if any of these options are expected to produce detectable $\gamma$-rays since the gas cloud into which the ejected mass is colliding is much less dense than the wind from a red giant (RG) \cite[]{Cheung2016}, and in novae with short recurrence times the amount of ejected mass is low \cite[]{Prikov1995,Yaron2005,Hillman2015,Hillman2016,Shara2018,Hillman2019}. Nevertheless, there are peculiar detections of $\gamma$-rays in some novae hosting a RD donor, as mentioned earlier \cite[]{Ackermann2014,Cheung2016,Martin2018}.
In symbiotic systems, where high energy $\gamma$-rays are plausible (as explained above), it is still not entirely clear what nuclear process is emitting them. The main interpretation of this high energy emission has been claimed to be due to hadronic particle acceleration in shocks \cite[]{Steinberg2020,Acciari2022,Hess2022}.  High energy protons, accelerated in the shock region may interact with other protons in the dense environment, giving rise to neutral pions ($\pi^{0}$) that then decay to high energy $\gamma$-rays. A proton-proton interaction will also produce charged pions ($\pi^{\pm}$) that will decay into high energy neutrinos. Modeling $\gamma$-ray emission from an astrophysical source with
a $\pi^{0}$ model thus inevitably predicts a high-energy neutrino flux
from the same source \cite[e.g.,][]{Stecker1970}. 
Therefore, if the high energy $\gamma$-ray emission has an hadronic origin, we expect the process to be accompanied by the production of neutrinos. 
This work aims to test the origin of the physical processes responsible for the $\gamma$-ray emission that is sometimes observed in nova eruptions. If the high-energy emission observed in these transients has an hadronic origin, it should be accompanied by a flux of neutrinos \cite[]{Razzaque2010,Metzger2016,Bednarek2022}. In this work we estimate the neutrino flux that might be associated with the recent eruption of RS Oph and thus we predict the number of events that, in principle,  could be detected during nova explosions by present and future neutrino telescopes.

In \S \ref{sec:telescopes} we specify the technical capabilities of each neutrino detector that we refer to in this work. \S \ref{sec:100GeV} specifies our methods of calculations for the different energy ranges of neutrino flux followed by our results in \S \ref{sec:results}. We discuss the implications of our results and compare the expected number of neutrino events with those derived in previous works in \S \ref{sec:discussion} and provide our conclusions in \S \ref{sec:conclusions}.

\section{Neutrino Telescopes}\label{sec:telescopes}

High-energy neutrinos interact with nucleons, producing secondary particles that travel faster than the speed of light in the sea or ice, inducing Cherenkov radiation inside the detector. The photons that are emitted by this process are detected by optical sensors that are deployed in the sea or ice (depending on the detector). In the following we briefly describe the basic characteristics of each telescope considered in this work.

\begin{figure}[ht]
\begin{center}
\includegraphics[width=0.5\textwidth]{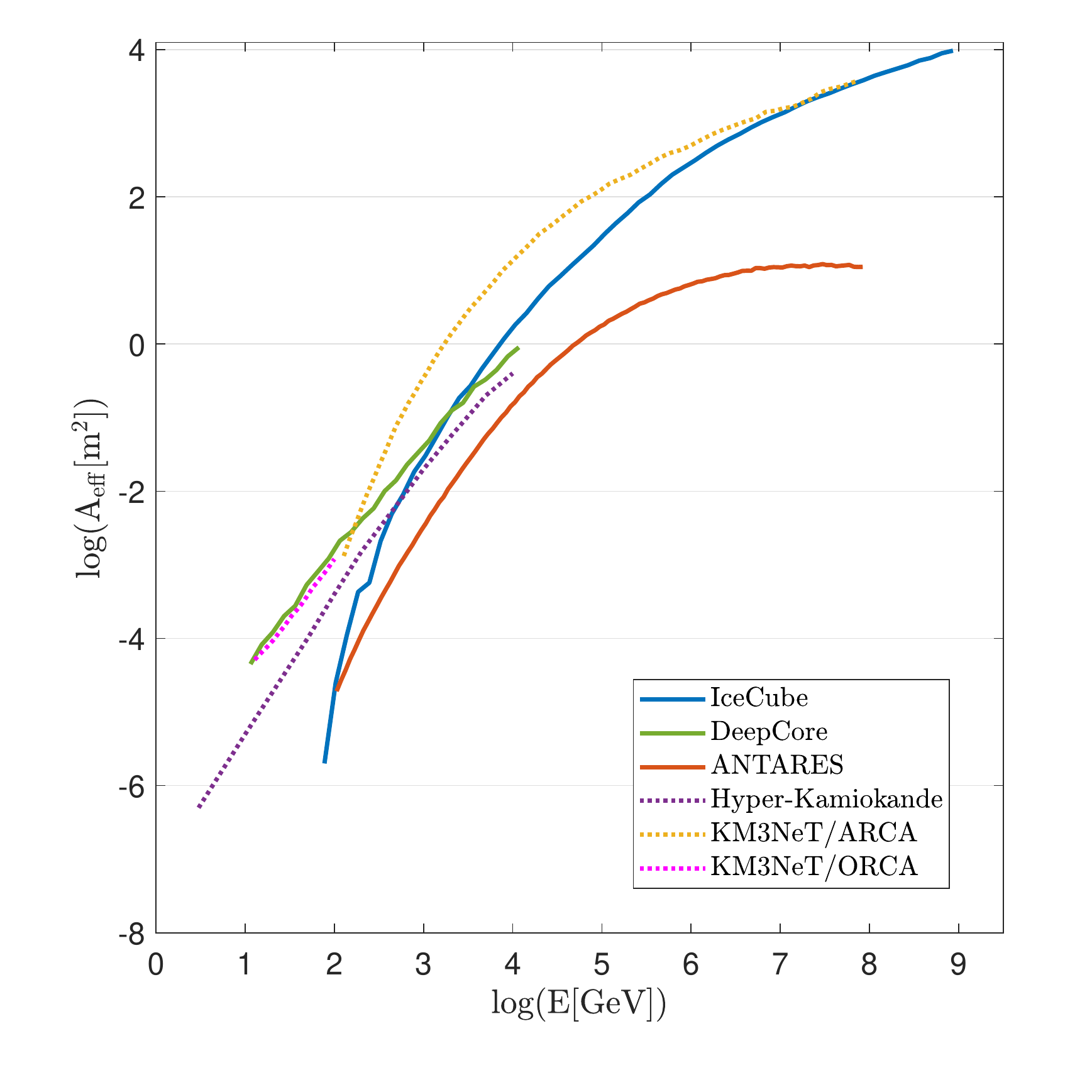}
\caption{The effective area vs. energy (on a log-log scale) for the six detectors, reproduced from: \cite{IcePoint} (IceCube); \cite{Zegarelli2022} (DeepCore and KM3MeT/ORCA); \cite{ANT} (ANTARES); \cite{detKM3} (KM3NeT/ARCA); and \cite{HyperKfig} (Hyper-K). Current telescopes (solid); future telescopes (dashed).} 
\label{fig:telescopes}
\end{center}
\end{figure}

\fix{IceCube and DeepCore} -- The IceCube high-energy neutrino telescope is a neutrino detector located at the geographic South Pole \cite[]{detIce} and covers a surface area of roughly 1 $\rm km^2$. 
Inside IceCube there is also the densely instrumented central DeepCore detector \cite[]{Ice2,Ice3}.
The module density in DeepCore is about five times greater than the rest of IceCube, which
allows for the much lower energy detection threshold of a few GeVs.
The IceCube detector has been collecting
data since 2006, and so far no neutrino event has been
associated with a nova eruption. The effective areas vs. neutrino energy for IceCube and DeepCore are shown in Figure \ref{fig:telescopes}.

\fix{ANTARES} -- The ANTARES neutrino detector is located in the Northern Hemisphere and is the only deep sea high energy neutrino telescope \cite[]{detAN} that exists to date. The telescope covers an area of about 0.1 $\rm km^2$ on the sea bed, at a depth of 2475 m, 40 km off the coast of Toulon, France. 
Figure \ref{fig:telescopes} shows the effective area of the ANTARES neutrino detector, with selection and reconstruction criteria optimized for the search of  point like sources, as a function of the neutrino energy \cite[]{ANT}.

\fix{KM3NeT} -- The KM3NeT detector \cite[]{detKM3} is the future generation of under water neutrino telescopes. 
KM3NeT will be comprised of the KM3NeT/ARCA which will consist of two building blocks each that will be deployed at a depth of 3500 m at a site 80 km South-East of Porto Palo di Capo Passero, Sicily, Italy, and of a third building block, called KM3NeT/ORCA, which will be located at a depth of 2200 m in a site close to ANTARES (Toulon), France.
KM3NeT/ARCA will have large spacings between adjacent strings in order to target astrophysical neutrinos at TeV energies.
The KM3NeT/ORCA will be sensitive to neutrinos down to energies of $\sim10$GeV thanks to the denser and compact array. Figure \ref{fig:telescopes} shows the effective areas of KM3NeT/ARCA and KM3NeT/ORCA as a function of the neutrino energy \cite[]{detKM3,Zegarelli2022}.

\fix{Hyper-Kamiokande (Hyper-K)} -- The Hyper-Kamiokande is a next generation under-water
Cherenkov detector with a sensitivity that is far beyond that of the Super-Kamiokande (Super-K) detector. The Hyper-K is designed to detect proton decays, atmospheric neutrinos, and neutrinos from astronomical origins. The baseline design of Hyper-K is based on the highly successful Super-K, taking full advantage of a well-proven technology \cite[]{HyperKfig}.
In Figure \ref{fig:telescopes} we show the effective area of Hyper-K as a function of the neutrino energy. As may be seen in this figure, the detector has good low energy performance, which should allow detection down to a few GeV.

\section{Expected neutrino fluxes}\label{sec:100GeV}

In this section, we derive the neutrino flux expected from RS Oph based on the assumption that the high energy photon emission is due to hadronic processes. Relativistic protons may produce $>$ GeV  $\gamma$-rays either by photo-meson production or inelastic nuclear collisions. In \cite{Bednarek2022} the authors show that p-p interactions are the most likely mechanism for pion production in novae. A possible mechanism that can produce the very high energy (VHE) photons that were detected by H.E.S.S. \cite[]{Hess2022} and MAGIC \cite[]{Acciari2022} may be the decay of neutral pions ($\pi^0$) produced through nuclear collisions of relativistic
protons. 
The same process that produces the neutral
pions, and subsequently the sub-TeV photons, would also generate
charged pions ($\pi^\pm$) that decay into neutrinos of similar energy. The following equation describes the three processes:
\begin{equation}
p+p \rightarrow \pi^0, \pi^+, \pi^- p, n, ...
\end{equation}

From this kind of interaction we expect almost the same number of $\pi^+$,$ \pi^0$ and $ \pi^-$ particles, due to isospin symmetry \cite[e.g.,][]{Povh2004}. $\pi^0$ particles decay into two $\gamma$-rays, having, in the pion rest frame, an energy equal to half of the $\pi^0$ mass as described below: 

\begin{equation}
\pi^0\rightarrow \gamma + \gamma
\end{equation}

On the other hand, the charged pions decay into neutrinos as follows: 
\begin{equation}
\pi^+ \rightarrow \mu^+ + \nu_{\mu} \rightarrow e^+ + \nu_e + \bar{\nu}_{\mu}+ \nu_{\mu}
\end{equation}

\begin{equation}
\pi^- \rightarrow \mu^- + \bar{\nu}_{\mu} \rightarrow e^- + \bar{\nu}_e + \nu_{\mu} + \bar{\nu}_{\mu}
\end{equation}
where $\nu_\mu$ and $\nu_e$ are the muon and electron neutrinos respectively.
Considering the relation between the photon flux and the neutrino flux given in Eq. 4 of  \cite{Razzaque2010} we derive that \cite[]{DiPalma2017}:

\begin{equation}\label{eq:pions}
\frac{dN_{\nu+\bar{\nu}}}{dE_{\nu}}=\frac{dN_{\gamma}}{dE_{\gamma}}
\end{equation}

and therefore
\begin{equation}
\int_{E_{\nu}^{\rm min}}^{E_{\nu}^{\rm max}}E_{\nu}\frac{dN_{\nu}}{dE_{\nu}}dE_{\nu}= \int_{E_{\gamma}^{\rm min}}^{E_{\gamma}^{\rm max}}E_{\gamma}\frac{dN_{\gamma}}{dE_{\gamma}}dE_{\gamma}
\end{equation}
where $ E_{\gamma}^{\rm min}$ ($E_{\nu}^{\rm min}$) and $E_{\gamma}^{\rm max}$ ($E_{\nu}^{\rm max}$) are the minimum and maximum photon (neutrino) energies respectively.

Today there is uncertainty about which theoretical model can give a good fit to a general functional form of the $\gamma$ ray flux (from novae) for any energy range. The established theory for proton acceleration follows a power law and,  in a simplified model, the photons follow the same power law. However, a single power law does not fit both the low and high-energy $\gamma$-rays. This may indicate that a single external shock cannot reproduce the observed spectrum at low and high energy. In a recent paper, \cite{Diesing2022} show that the spectrum at low and high energy can be reproduced if multiple shocks are considered. The authors perform detailed, multi-zone modeling of RS Ophiuchi’s 2021 outburst including a self-consistent prescription for particle acceleration and magnetic field amplification. We approximate the functional form of $\gamma$-ray flux (from novae) for any energy range, that approximately can reproduce this model, with a  log-parabola function. We would like to emphasize that this is a purely phenomenological fit.
We use the following log-parabola spectral function for the entire range range ($>1$ GeV):

\begin{equation}\label{eq:lowE}
\frac{dN_{\gamma}}{dE_{\gamma}}=N_0\frac{E}{E_0}^{-\alpha-\beta \rm{ln}(E/E_0)}
\end{equation}

where $E_0\sim 1$ GeV is the reference energy,  $N_0\sim 4\times 10^{-4} {\rm TeV}^{-1} {\rm cm}^{-2} {\rm s}^{-1}$ is the amplitude at the reference energy (that we considered averaged over the 5 nights),  $\alpha \sim 2$ is the spectral index and $\beta \sim 0.15$ is the curvature. All these data are taken from  table S3 of \cite{Hess2022}. This data is used to estimate the expected number of neutrinos or all six detectors.

Following the line of the work by \cite{Alvarez2002}, \cite{Guetta2003} and \cite{DiPalma2017}, we compute the high energy neutrino flux at Earth and estimate the number of events that may be detected by the telescopes described in \S\ref{sec:telescopes}.

The total number of expected astrophysical events during an exposure time $T$ of a neutrino telescope
is given by:
\begin{equation}\label{eq:integral}
N=\int_{E_1}^{E_2} T \frac{dN_{\nu}}{dE_{\nu}}A(E_{\nu})dE_{\nu}
\end{equation}
where $\frac{dN_{\nu}}{dE_{\nu}}$ can be derived from $\frac{dN_{\gamma}}{dE_{\gamma}}$ according to Equation \ref{eq:pions} for the given spectrum in the energy range $E_1-E_2$  (and used in Equation \ref{eq:TeVspec}) and $A(E_{\nu})$ is the effective area of the considered neutrino telescope, as a function of the neutrino energy $E_{\nu}$ as shown in Figure \ref{fig:telescopes}.  

The effective area of a detector may depend on the declination of the observed celestial object. The declination of RS Oph is $-06^0 4^{'} 28.5^{''}$, and we use the corresponding effective areas of the relevant detectors where applicable.

\section{Results of calculations}\label{sec:results} 

In this section we show the results of our calculations regarding the 2021 RS Oph eruption for the six detectors. We also apply our analysis to additional novae (elaborated in \S \ref{sec:intro}) that were detected in $>$1GeV by Fermi-LAT. 

\subsection{RS Oph - High energy ($>100$ GeV)}\label{sec:high_energy}
For the high energy regime we use Equations \ref{eq:integral}$-$ \ref{eq:lowE}  and the data 
analysis from \cite{Hess2022} to calculate estimates of the total number of neutrinos expected to have been detected from the latest RS Oph eruption by IceCube, ANTARES and KM3NeT/ARCA. For each detector, we calculate the average number of events per hour over the five exposure epochs. According to \cite{Hess2022} the source was observed in eruption for $\sim30$ days. In order to obtain an upper limit of the total number of expected neutrino events, we multiply the average number that we obtain by 30 days. 
We find the total expected number of neutrino events  for IceCube, ANTARES and KM3NeT/ARCA to be $\sim0.5\times 10^{-3}$, $\sim1.5\times10^{-4}$ and $\sim0.8\times10^{-2}$ respectively. The difference of the log-parabola parameters for the five nights, affect the neutrino flux by a factor of three.

\subsection{RS Oph - Low energy ($1-100$ GeV)}\label{sec:low_energy}
Next we calculate the expected flux for the case that the neutrinos could have originated in $\pi$ decays resulting in lower energies. As for the high energy regime, we use Equations \ref{eq:integral}$-$ \ref{eq:lowE} and data from \cite{Hess2022} as described in \S\ref{sec:100GeV}. We obtain $\sim0.014$, $\sim0.06$ and $\sim 0.046$ for Hyper-K, DeepCore and KM3NeT/ORCA respectively. 
The difference of the log-parabola parameters for the five nights, affect the neutrino flux by a factor of three.

\begin{figure}[ht]
\begin{center}
\includegraphics[width=0.5\textwidth]{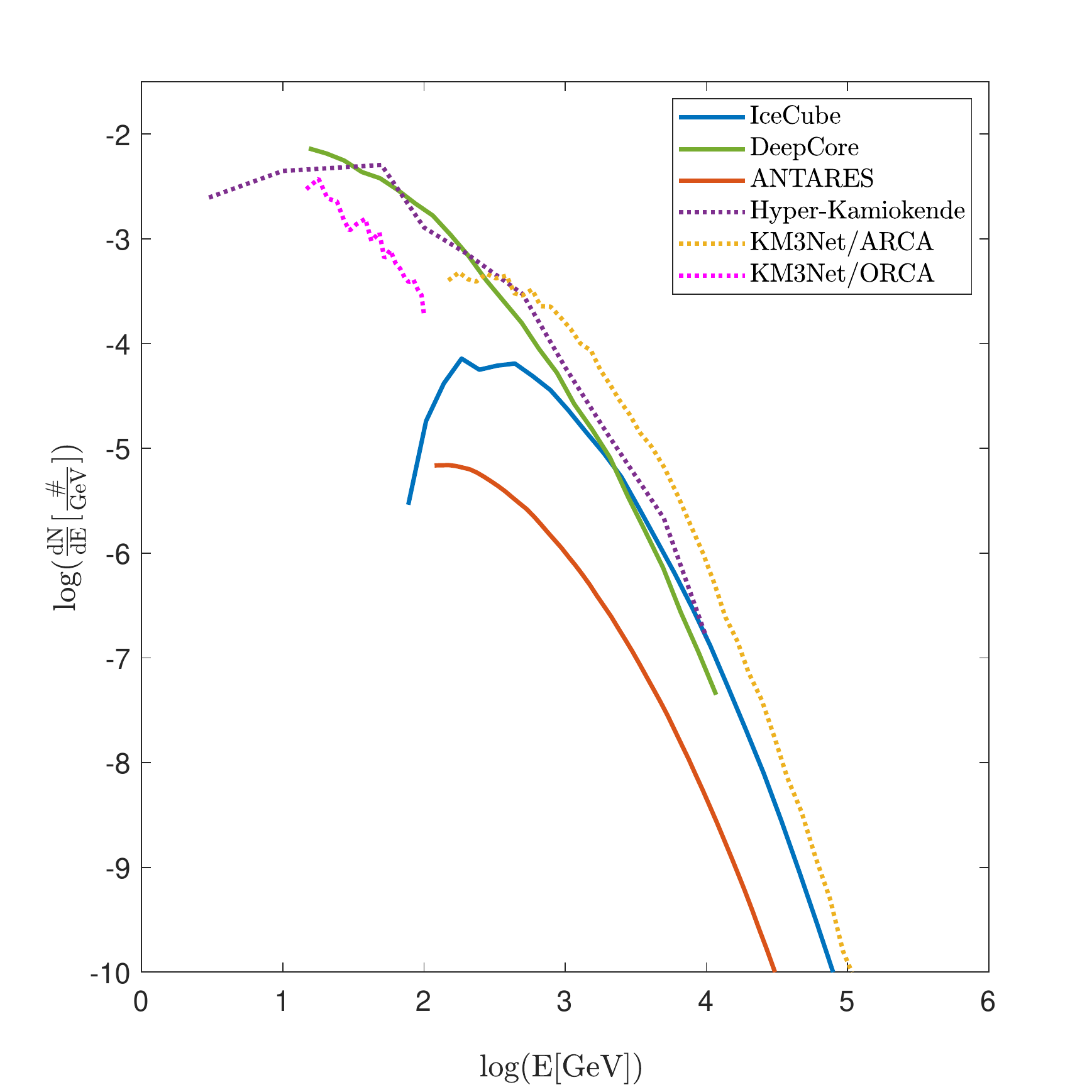}
\caption{Number of neutrinos, from RS Oph, expected to be detected by the six detectors, as a function of the netrino energy. The line colors and types are as in Figure \ref{fig:telescopes} for convenience. In the hypothetical case where we examine placing RS Oph at about one tenth of its actual distance (0.2 kpc) the curves would be shifted up by two orders of magnitude. }
\label{fig:dNdE}
\end{center}
\end{figure}

In Figure \ref{fig:dNdE} we show the distribution of the number of expected neutrinos from RS Oph, over the entire energy regime, for the six detectors. We used the log-parabola function form to produce this distribution. This figure shows the the sensitivities of current and future detectors as a function of neutrino energy, illustrating how the spectral
shape of RS Oph affects its detection prospects.
\subsection{Expected atmospheric events}

We have presented quantitative estimates of future neutrino detection from novae with current (IceCube-DeepCore) and under construction (KM3NeT/ORCA and Hyper-Kamiokande) neutrino telescopes. However, at multi-GeV energies the atmospheric background severely limits the identification of cosmic signals. The main component for the background is the flux of atmospheric neutrinos, which is caused by the interaction of cosmic rays, high energy protons and nuclei, with particles in the Earth's atmosphere. Decays of charged pions and kaons produced in cosmic ray interactions generate a flux of atmospheric neutrinos and muons. In order to decrease the effect of the background, the search of neutrinos with energies in the multi-GeV range from novae should be performed only for upward going neutrinos. Indeed, Earth-filtered events allow to reduce the atmospheric muon background significantly. Moreover, only events due to $\nu_{\mu}$  charged current (CC) interactions should be considered. The muons that originate in such interactions indeed lead to a long track that allows to define the direction of an incoming neutrino with good accuracy, pointing back to the source. 

An approximate estimate of the background events is given in \cite{Metzger2016}. They estimate the number of background events to be $\sim 1$ neutrino over a two week duration (which is a typical time span for $\gamma$-ray emission in novae) for DeepCore-IceCube. Following their work we estimate 2.2 neutrinos as background events for an epoch of 30 days, corresponding to the observation time of RS Oph.  

\subsection{Additional novae}\label{sec:additional_novae}

We now consider the Fermi-LAT detections of the six novae specified in \S \ref{sec:intro}.The flux of these novae can be fitted by a power-law that expresses the number of photons per unit energy interval, time, and surface area. This can be written as:

\begin{equation}
\label{eq:TeVspec}
\frac{dN_{\gamma}}{dE_{\gamma}}=N_0\left(\frac{E}{E_0}\right)^{-\Gamma}
\end{equation}

where $N_0$, $E_0$ and $\Gamma$ are the amplitude at reference energy, reference energy and spectral index respectively, computed from observations.
We take the photon fluxes from \cite{Ackermann2014} and \cite{Cheung2016} and use them in Equation \ref{eq:TeVspec} and \ref{eq:integral} to determine the expected number of neutrino events, while assuming, as before, that the energy emitted in neutrinos is of the same order as the energy emitted in photons. We used $E_0=1$ GeV as the reference energy and $\Gamma=2.1$ as the spectral index. The values for exposure time (T) and amplitude at reference energy ($N_0$) (extracted from \cite{Ackermann2014} and \cite{Cheung2016}) are specified in Table \ref{tab:six_novae} as well as our resulting number of expected neutrino detections by each of the low energy telescopes up to 100 GeV.

Our results predict, for the six novae, substantially smaller numbers of expected events relative to RS Oph, and within the six novae, we expect a higher detection rate for the SymN V407 Cyg relative to the five CNe. We note that none of the low energy telescopes yield a feasible number of expected events for any of these novae.

\begin{table*}[]
	\begin{center}
		\begin{tabular}{c|ccc|ccc|c}
{ }&{D[kpc]}&{$N_0$} & {T[days]} & {$N_{\nu}^{\rm DeepCore}$} & {$N_{\nu}^{\rm Hyper-K}$}& {$N_{\nu}^{\rm ORCA}$} &{$E_{\nu}^{\rm TOT}$} \T\B\\
\hline	
{V339 Del}& {4.2}&{5.0}&{27} &{0.013}&{0.0026} &{0.009}&{6.0}\T\B\\
{V959 Mon}&{3.6}&{7.0}&{22} & {0.015}&{0.0030} &{0.011}&{7.1}\T\B\\
{V1324 Sco}&{4.5}&{10.0}&{17} & {0.016}&{0.0033} &{0.012}&{13}\T\B\\
{V407 Cyg}&{2.7}&{10.0}&{22} & {0.021}&{0.0043} &{0.015}&{6.1}\T\B\\
{V1369 Cen}&{2.5}&{2.5}&{18} & {0.004}&{0.0009} &{0.003}&{3.0}\T\B\\
{V5568 Sgr}&{2.0}&{1.0}&{47} & {0.005} &{0.0009} &{0.003}&{1.2}\T\B\\
\hline\hline	
{RS Oph}&{2.3}&{7.1}&{30} & {0.060} &{0.0140} &{0.046}&{20}\T\B\\
\end{tabular}
\caption{Summary of six novae. D and T are the distance to the system and the exposure time and $N_0$ is given in units of $10^{-11}\rm erg^{-1}cm^{-2}s^{-1}$  \cite[]{Ackermann2014,Cheung2016}. Columns $3-5$ are the derived expected number of neutrinos for the three low energy detectors. In the last column we give the total energy emitted in neutrinos in units of $10^{41}$ erg. The data of RS Oph for the log-parabola function form are included for comparison.}\label{tab:six_novae}
\end{center}
\end{table*}

Additionally, we extrapolate the above calculation to predict the number of neutrino events for the hypothetical case that those novae may emit in the high energy range ($>100$ GeV). We accomplish this by extending the Fermi-LAT photon flux to higher energies. In order to get the high energy neutrino flux, \cite{Razzaque2010} extrapolate the low energy flux to high energies by using the low energy spectral index. We follow this procedure for the six novae and obtain the hypothetical high values of $\sim0.2-3.5$ neutrino events. 
However, we note that RS Oph is a particularly luminous nova, likely characterized by an unusually fast shock velocity, large ambient density, and therefore a high maximum proton energy. Less luminous novae may \textit{not produce 100 GeV emission at all.} The implication of this will be elaborated in \S \ref{sec:discussion}.

In order to test if this may be considered a realistic number of events, we then use this extrapolation method to calculate an expected number of neutrinos from RS Oph in the high energy range, and obtain $\sim12$ neutrinos. After comparing this number with the results obtained in \S \ref{sec:high_energy} where we use the high-energy photon observations of RS Oph to obtain a realistic estimate of the number of neutrinos possibly emitted by RS Oph, we find that the "predicted" number is about a factor  $\sim 5\times10^3$ greater than that derived by taking the geometric mean of the three observed values (\S \ref{sec:high_energy}). This implies that the RS Oph spectrum cannot be represented by a a single power law indicating that the observed $\gamma$-rays cannot arise from a single, external shock \cite{Diesing2022}.

\section{Discussion}\label{sec:discussion}

In this section we discuss the fact the RS Oph showed a much higher $\gamma$-ray and, possibly, neutrino flux in high energy than any other nova to date. This cannot be explained by the flux fading with distance since, as shown in Table \ref{tab:six_novae}, of the six novae some are closer than RS Oph and some are further but all have a much lower number of expected neutrino events.
Being more energetic means the basic system parameters (e.g., WD mass, donor type or mass or evolutionary stage, separation, accretion rate, kinetic energy etc.) would have to be different. However, understanding what system parameters may produce sufficiently energetic interactions is not straightforward. For instance, let us consider the extreme, rapidly recurring nova, M31N 2008-12a \cite[]{Darnley2016} that erupts every year. It should be producing multiple mass shells that expand away from the WD, and they would not all be expanding at the \textit{exact} same velocity, inevitably leading to collisions between different shells. In-homogeneity in the ejecta can form clumping which can lead to collisions as well. This interpretation can mislead to the simplistic conclusion that a system with a shorter recurrence period should be the place to look for highly energetic shocks. However, the amount of mass ejected in a nova decreases with decreasing time between eruptions. This means that being a recurrent nova is not necessarily the only requirement. RS Oph, being a SymRN, is embedded in the dense wind coming from its companion, so the nova eruption sends the ejected shell hurdling into it, which is the source of the GeV radiation. This being the case, perhaps we should expect to find this range of energy in all SymNe? \cite{Ackermann2014} and \cite{Razzaque2010} have investigated the SymN V407 Cyg and found, for the relevant energy range, lower fluxes  than found for RS Oph (based on kinetic energy considerations). We find similar results here for the low energy detectors.

What is the cause of this stark difference between these two SymNe? Both RS Oph and V407 Cyg host massive WDs ($\sim1.3$ and $\sim1.2M_\odot$ respectively , both have giant donors of about $1M_\odot$, but their orbital periods are very different --- $\sim1.5$ and $\sim43$ years respectively \cite[]{Munari1990,Pan2015,Shara2018,Hillman2021}. Additionally, RS Oph erupts every $\sim15$ years while V407 Cyg has one recorded eruption. This implies that the WD in RS Oph should be enduring a higher accretion rate than the WD in V407 Cyg. It is therefore tempting to make the following speculative suggestion: this high accretion rate implies that the donor's wind density at the WD is lower for V407 Cyg than for RS Oph, and the lower wind density may lead to less interaction between the nova ejecta and the giant's wind, thus, resulting in a lower $\gamma$-ray flux.

Stemming from this, we turn to investigate if high energy $\gamma$-ray emission was detected in other SymRNe. V745 Sco is a SymRNe with a recurrence time similar to that of RS Oph ($\sim20$ years). \cite{Delgado2019} report a factor of $\sim25$ between the $\gamma$-ray fluxes of the two nova which is compatible with the factor of order in their distance. This indicates that high energy ($>$1GeV) $\gamma$-rays from nova eruptions, should be expected only in SymRNe.

We note that we may have been systematically underestimating the neutrino emission from RS Oph due to the fact that we have not considered absorption of GeV$-$TeV photons from the surrounding environment. The neutrino flux may be larger than what found from the high energy photon flux \cite[]{lamastra16}. We remark that the connection between neutrinos and gamma rays is highly dependent
on the astrophysical environment. It may be that the neutrino source is obscured in the $\sim$ GeV$-$TeV range but bright in high energy neutrinos \cite[]{Fasano2022}. The calculation of this effect is not straightforward, since it involves modelling of the environment, including possible ancient shells that have expanded parsecs away from the source. However it is worth noting that substantial gamma-ray absorption is unlikely for RS Oph \cite[]{Hess2022,Diesing2022}.

It has been suggested that the expected signal event rate may be increased by combining search among low and high energy neutrino detectors, i.e., KM3NeT/ORCA + KM3NeT/ARCA and
DeepCore + IceCube \cite[]{Zegarelli2022}.
Another option that can greatly increase the signal detection is summing the contribution of many novae (stacking). However, the same holds for the atmospheric background, such that complex stacking techniques are required in order to obtain a significant detection level. (See \cite{Zegarelli2022} for a detailed description of this procedure).

\section{Conclusions}\label{sec:conclusions}
In this work we have estimated the number of neutrino events which are expected to be detected from novae by present and future neutrinos telescopes.  Our approach is calculating the number of expected neutrinos \textit{directly} from the observed high energy photon flux, in a model- independent way. We obtain a number of interesting results, specified below: 

\begin{enumerate}

\item
Given the current telescope sensitivity, neutrino emission is unlikely to be detected by novae and therefore cannot be used to confirm that the emission is hadronic. However other studies (see \cite{Hess2022,Acciari2022,Diesing2022}) show that the leptonic emission is subdominant \cite[]{Diesing2022} and that the hadronic interpretation is favoured. 

\item
RS Oph is observationally unique:  it is the only nova to date that has been observed in both GeV and TeV. Other, typical, novae may not exert sufficient kinetic energy to accelerate protons to the high energy regime. This includes the six other novae that we have analyzed in this paper, then implying that extrapolation from an observed energy regime to a non-observed regime may be entirely misleading.

\item\label{item:Gen2}
Our predictions for the number of neutrino events, both for the high and low energy ranges, are quite low. For the IceCube-DeepCore detections we estimate that a nova eruption, similar to RS Oph, must be at a distance not larger than $\sim 0.2$ kpc in order to obtain a $\sim3\sigma$ detection above the background. Note that the $3\sigma$ confidence limits has been calculated following the prescriptions for small numbers of events developed by 
\cite[]{Gehrels1986}. 
 
All the novae in the sample explored in this work are characterized by distances of $\sim$2 kpc or greater. However, the expected improvement for the IceCube-Gen2 \cite[]{Aartsen2021} detector would indicate that its detection capability could increase by about an order of magnitude.  In such a case, the detection distance may increase up to $\sim$0.5 kpc.

\item\label{item:4} 
The global nova rate in the Milky Way has been measured many times by several authors over the past decades (see \cite{DellaValle2020} for a summary). Currently, the frequency of occurrence of novae within the Galaxy is typified by a factor of two of uncertainty. Today it is commonly believed that its value is between 20 \cite[]{DellaValle1994} and 50 novae/year \cite[]{Shafter2017}.  Our location in the Galaxy, in the outskirts of the galactic disk together with the requirement of distances less than $\sim $ 1 kpc, limit our interest only to the disk nova component. We can assume that RS Oph-like (i.e. SymRN) events are the best candidates to produce high-energy
gamma rays and observable neutrino fluxes. The frequency of occurrence of novae in symbiotic systems is not well known, but could well be comparable to that of CNe \cite[]{Munari92}.
Following \cite{DellaValle1993} and using recent values for nova rates, we compute an upper limit for the nova eruption density in the disk of $\sim 1 \times 10^{-9}$ pc$^{-3}$ yr$^{-1}$. Given a ratio  RNe/CNe $\sim 0.3$ \cite[]{DellaValle2020,Pagnotta14,DVlivio96} we obtain a lower limit for future neutrinos detections from nova explosions in the Milky Way of $\sim 1$ event in about 30 years. Very recently, \cite[]{Mandel2022} calculated the specific nova rate for a spiral galaxy like M51 to be a factor about three larger than previously estimated.  Then we can set, as a realistic upper limit to the neutrino detection, a rate of one event in about 10 years or so. 
 
 
\item
With the goal being to characterize the system parameters that may produce these high energy photons, we investigated the basic parameters of a number of systems that belong to different system types --- CNe, RNe, SymNe and SymRNe --- and found that the most likely type of system to produce detectable emission in the $>$100 GeV regime would be novae that occur in SymRNe systems. The main reason for this being that these systems have a high accretion rate --- resulting from a dense wind, which is required for the collisions that are responsible for the high energy photons.


\end{enumerate}

\section*{Acknowledgements}
The support from the Authority for Research $\&$ Development and the chairman of the Department of Physics in Ariel University are gratefully acknowledged. Massimo Della Valle
thanks the University of Ariel for their hospitality during the visit and creative atmosphere

\bibliography{main}{}
\bibliographystyle{aasjournal}

\end{document}